\documentclass[usenatbib]{mnras}
\usepackage{widetext}
\usepackage{hyperref}
\usepackage{mathtools, cuted}

\begin{document}
\title[Cosmological constraints from thermal Sunyaev Zeldovich power spectrum revisited]{Cosmological constraints from thermal Sunyaev Zeldovich power spectrum revisited}

\author[B. Horowitz and U. Seljak]{B. Horowitz$^{1}$\thanks{E-mail: bhorowitz@berkeley.edu}, U. Seljak$^{1,2}$\thanks{E-mail: useljak@berkeley.edu}\\
$^{1}$Department of Physics, University of California at Berkeley, Berkeley, CA 94705\\
$^{2}$Lawrence Berkeley National Laboratory, Berkeley, CA 94720}

\maketitle

\begin{abstract}

Thermal Sunyaev-Zeldovich (tSZ) power spectrum is one of the most sensitive methods to constrain cosmological parameters, scaling as the amplitude $\sigma_8^8$. It is determined by the integral over the halo mass function multiplied by the total pressure content of clusters, and further convolved by the cluster gas pressure profile. It has been shown that various feedback effects can change significantly the pressure profile, possibly even pushing the gas out to the virial radius and beyond, strongly affecting the tSZ power spectrum at high $l$. Energetics arguments and SZ-halo mass scaling relations suggest feedback is unlikely to significantly change the total pressure content, making low $l$ tSZ power spectrum more robust against feedback effects. Furthermore, the separation between the cosmic infrared background (CIB) and tSZ is more reliable at low $l$. Low $l$ modes are however probing very small volumes, giving rise to very large non-gaussian sampling variance errors. By computing the trispectrum contribution we identify $90<l<350$ as the minimum variance scale where the combined error is minimized. We find constraints on $\sigma_8$ by marginalizing over the feedback nuisance parameter, obtaining $\sigma_8 =0.820^{+0.021}_{-0.009}\left(\Omega_m/0.31\right)^{0.4}$ when fixing other parameters to Planck cosmology values.  Our results suggest it is possible to obtain competitive cosmological constraints from tSZ without cluster redshift information, and that the current tSZ power spectrum shows no evidence for a low amplitude of $\sigma_8$.

\end{abstract}


\section{Introduction}

The thermal Sunyaev Zeldovich (tSZ) effect is a secondary anisotropy of the cosmic microwave background (CMB), where CMB photons inverse Compton scatter off of energetic electrons that lie along the line of site between us and the surface of last scattering \cite[]{sunyaev1970small}. Its amplitude $Y$ is determined by the projected gas pressure along the 
line of sight. The tSZ effect has been used for over a decade to study individual clusters ~\cite[]{reese2002determining}. The pressure can be expressed as a product of density and temperature, and in 
a virialized system the 
latter scales roughly as $M_{\rm vir}^{2/3}$, where $M_{\rm vir}$ is the halo virial mass. 
Integrating the tSZ signal across the cluster gives the scaling of $Y_{\rm vir} \propto M_{\rm vir}^{5/3}$. 
In recent years the multi wavelength, high angular resolution, large-array surveys has allowed the measurement of its power spectrum over a large range of scales ~\cite[]{aghanim2015planck}~\cite[]{story2013measurement}. 

The tSZ power spectrum has been advocated as a
strong probe of cosmology, ranging from constraining $\Lambda$CDM cosmological parameters ~\cite[]{komatsu2002sunyaev}, to primordial non-gaussianities and massive neutrinos ~\cite[]{hill2013cosmology}. Its main advantage when compared to cluster abundance method is that one does not need to measure the cluster redshifts, or their virial halo mass. Instead, tSZ power spectrum probes an integral over the cluster halo abundance as a function of redshift and halo mass, multiplied by the total pressure content of clusters, and further convolved by the cluster gas pressure profile. 
tSZ power spectrum is sensitive to different halo masses and redshifts as a function of angular moment $l$. 
However, it is still a projection and as such it is 
difficult to disentangle the different redshifts and/or 
halo masses. 

The tSZ power spectrum is sensitive not only to cosmological parameters, but also to nonlinear gas physics found in the intra-cluster medium ~\cite[]{komatsu1999sunyaev, battaglia2010simulations}. The gas distribution depends on the dark matter gravitational potential well, stellar formation, AGN feedback, supernovae, and radiative cooling. In practice,  simulated pressure profiles in different simulations often differ with one another, and the resulting power spectra differ as well, specially at high $l$ where the change of profile matters more \cite[]{mccarthy2014thermal}. Direct observation of the pressure profiles is a promising approach ~\cite[]{arnaud2010universal}, but is limited to observed massive halos where the effects of AGN and other feedback mechanism are less pronounced. 

In this paper we revisit tSZ power spectrum as a probe of 
cosmology.
We take advantage of the fact that while the gas pressure profiles are very dependent on the detailed physical modeling inside the clusters, the total pressure content integrated over the cluster is a 
lot less model dependent. This is because while the feedback models can push the gas around, they cannot 
easily inject enough energy to change its total thermal content. This is confirmed in tSZ power spectrum simulations \cite[]{mccarthy2014thermal}, which are relatively unaffected at low $l$. Another argument are the scaling relations of Planck \cite[]{2013A&A...557A..52P} and by Greco et al.\cite[]{greco2015stacked}, where the simple $Y_{\rm vir} \propto M_{\rm vir}^{5/3}$ scaling holds over a large range of halo masses. A weak lensing
calibration of this scaling relation has recently been provided in \cite[]{wang2016weak}. 
The scaling works because the Planck beam is very large and encompasses the entire cluster tSZ effect, 
but it also suggests that there is no significant change in the thermal gas content that would break the scaling. 
We can thus 
side-step the complex gas dynamics by focusing on large scales where the gas profile is less important, 
and only the total pressure cluster content contributes. In this paper we will 
use a simple one-parameter model for AGN and supernova feedback, similar to the model used by for CMB weak lensing statistics~\cite[]{mohammed2014baryonic}. Various simulations show that gas will be expelled in less massive halos ~\cite[]{read2005mass},~\cite[]{pontzen2012supernova}, and forced to outer reaches beyond the viral radius. This parameter has the effect of suppressing the high $l$ contribution from clusters below a given critical mass ($M_{crit}$). In effect, the galaxy is ``puffed up'' due to the feedback effects. However, as stated above, in our model we preserve the total gas pressure content, so at sufficiently large scales
(low $l$) there is no effect. 

In section II we quickly overview the tSZ power spectrum calculations, its scaling relations, its covariance matrix model, and the feedback modeling. In section III we perform a likihood analysis over the $\sigma_8$, $M_\textrm{crit}$ parameter space and marginalize over $M_\textrm{crit}$ to find constraints on $\sigma_8$. In section IV we compare our constraints with other techniques and discuss the outlook for further method improvements. 

For all our analysis, we use the Planck TT,TE,EE+lowP+lensing+BAO fiducial cosmology for all our calculations; $\Omega_m = 0.3089$, $\Omega_\Lambda = 0.6911$, and $h = 67.74{\rm km/s/Mpc}$. We use for data the NILC - MILCA F/L cross-power spectrum after foreground subtraction as described in ~\cite{aghanim2015planck} and the ACT value at high $l$ from ~\cite{hill2014atacama}.

\label{sec:intro}



\section{Thermal SZ Power Spectrum}
\label{sec:tSZPS}

The full analytical description of the tSZ power spectrum and its dependencies are well covered in existing literature~\cite[]{komatsu2002sunyaev,hill2013cosmology}, and here we simply summarize the necessary results. We present our work in terms of the general Compton y-parameter and the results can be multiplied by the necessary $g_\nu$ factor for given frequency bands. Where temperature is referenced, we take $\nu = 146$ GHz where $g_\nu^2 = 1$. Like other tSZ studies, we ignore relativistic corrections to the tSZ power spectra as they primarily effect the most massive halos ($> 10^{15} M$) which primarily effect low $l$ modes outside of our range of interest~\cite[]{nozawa2005relativistic}.

The tSZ effect results in a frequency-dependent shift in the CMB temperature observed in the direction of a dense collection of hot electrons, such as a galaxy cluster. Utilizing the halo-model, we can write our power spectrum as a superposition of the one-halo and two halo terms:
\begin{equation}
C_l^{tot} = C_l^{1-halo}  + C_l^{2-halo} 
\end{equation}
The one halo term, in the flat sky limit ($l\gg 1$) is
\begin{equation}C_l^{1-halo} = \int_{z_{low}}^{z_{max}} dz \frac{d^2V}{d\Omega dz} \int_{M_{min}}^{M_{max}} dM \frac{dn(M,z)}{dM} |\tilde{y}_l(M,z)|^2\end{equation}
where $\tilde{y}_l$ is the Fourier transform (in the Limber approximation, see ~\cite[]{hill2013cosmology} for discussion of the validity of this approximation) of the pressure profile given by 
\begin{equation}\tilde{y}_l(M,z) = \frac{4 \pi r_s}{l_s^2} \int dx \frac{\sin{(l+ 1/2)x/l_s}}{(l+ 1/2)x/l_s}y_{3d}(x,M,z)x^2.\end{equation}
In these equations, $r_s$ is the characteristic scale radius, $r_{200,c}$, of the $y_{3d}$ profile and $l_s$ is the associated multipole moment. 

The two halo term has an additional functional dependence on the bias, $b(M,z)$, and the linear matter power spectrum, $P_{lin}$. 
We use the bias fitting function of ~\cite[]{tinker2010bias}, and a matter power spectrum generated by CAMB.
\section*{{}}
\vspace{-50pt}
\begin{widetext}
\begin{equation}
C_l^{2-halo} =  \int_{z_{low}}^{z_{max}} dz \frac{d^2V}{d\Omega dz} \left[ \int_{M_{min}}^{M_{max}} dM \frac{dn(M,z)}{dM} |y_l(M,z)| b(M,z) \right]^2 P_{lin} \left(\frac{l+1/2}{\chi(z)},z\right) \end{equation}
\end{widetext}
Following the Planck tSZ collaboration,~\cite[]{aghanim2015planck} we use the standard pressure profile of~\cite[]{arnaud2010universal}, the concentration parametrization of ~\cite[]{duffy2008dark}, and the mass function of ~\cite[]{tinker2008mf}. For our integration range we use the $M_{low} = 5 \times 10^{11} M_\odot$, $M_{high}=5 \times 10^{15}M_\odot$, $z_{low} = 0.0001$, and $z_{high} = 5.0$. 

We find our power spectra scales as roughly $C_l^{\textrm{max}} \propto (\Omega_b h^2)^2 \Omega_m^3\sigma_8^8$, consistent with results found in ~\cite[]{trac2011templates,shaw2010impact,aghanim2015planck}. Note that while the overall amplitude is directly proportional to $\sigma_8$, a slight shift is created when $\Omega_b$ is varied due to its complex relation with the linear power spectrum which enters directly in the 2-halo term as well as indirectly though conversion relations between characteristic mass and radii of halos. However, varying $\Omega_m$  within the range of cosmological interest only effects the amplitude of the tSZ powerspectrum \cite[]{komatsu2002sunyaev}. We quote our results as constraints on $\sigma_8\left(\Omega_m/0.31\right)^{0.4}$ to capture this degeneracy in the amplitude between $\sigma_8$ and $\Omega_m$.


\subsubsection{Comparison to Simulations}
\label{sec:sim}

\begin{figure}
   \includegraphics[clip, width=0.5\textwidth]{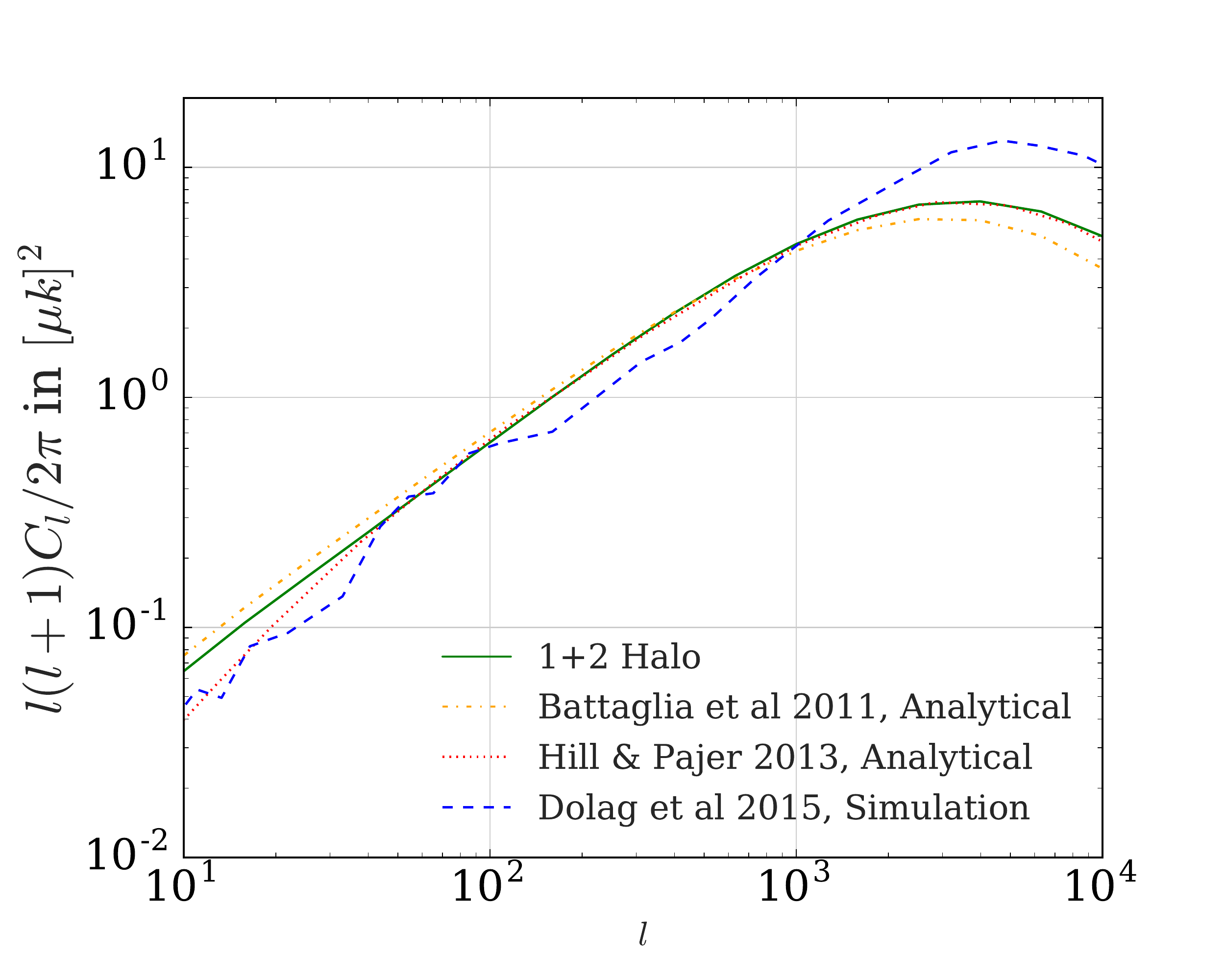}
    \caption{A comparison of our model (solid green) at 146 GHz, simulations of \protect\cite[]{dolag2015sz} (blue dotted), and the analytic model of Hill and Pajer (red dotted) \protect\cite[]{hill2013cosmology} and Battaglia et al (yellow dot-dashed) \protect\cite[]{battaglia2012cluster} which use different pressure profiles than our model. We have scaled all power spectra to the fiducial cosmology with $\sigma_8 = 0.815$}
    \label{fig1}
\end{figure}

Comparisons to simulations are not completely straight-forward due to low simulated volume used in hydrodynamic simulations, 
which induce large variance in tSZ power spectrum, which 
may differ from a global ensemble average. The best comparison so far has been done in \cite[]{battaglia2010simulations}, where this 
problem has been circumvented by inserting analytic tSZ profiles
directly into the simulations. They find good agreement, 
suggesting that the analytic models can be well calibrated against simulations. This was discussed further by \cite[]{hill2013cosmology}. Our model comparison against theirs
is shown in Figure \ref{fig1}.

We perform another comparison against 
recent tSZ power spectrum 
using Magneticum Pathfinder Simulation, presented in ~\cite[]{dolag2015sz}. We also compare our model 
to theirs in Figure \ref{fig1}. Overall there is good agreement, but there are fluctuations in the simulated spectrum at $l<1000$, which may 
be caused by the small simulated volume. At high $l$ our model, which 
includes feedback effects, differs strongly from \cite{dolag2015sz} which is also true of the other analytical models as discussed in their work. The offset between the analytical models in the range of interest $100<l<600$ corresponds with a 4\% change in the value of $\sigma_8$. More detailed comparisons of the simulations and various possible analytically models can be found in \cite{mccarthy2014thermal,dolag2015sz,battaglia2012cluster}.

\subsubsection{Comparison to Weak Lensing Studies}
\label{sec:weaklens}

In a recent work \cite[]{wang2016weak}, a relation between the cluster mass ($M_{500}$) and the gas pressure profile, which they denote as the integrated Compton $Y_{500}$ parameter, has been presented. It is instructive to compare our pressure profile with the results of this work. 

Their study used weak lensing to measure the mean pressure profiles of a number of Locally Bright Galaxies over a very broad mass range ($10^{12.5}M_\odot$ to $10^{14.5}M_\odot$). They find

\begin{equation} \tilde{Y}_{500} = Y_M \left( \frac{M_{500}}{10^{13.5}M_\odot} \right)^{\alpha_m},
\end{equation}

with best fit parameters $Y_M = 2.31^{+0.36}_{-0.38} \times 10^{-5}$ and $\alpha_m = 1.61^{+0.14}_{-0.18}$.

Using our pressure profile and cluster masses, and integrating over the total profile of the cluster, we found a similar fit, with $Y_M = 2.26 \times 10^{-5}$ and $\alpha_m = 1.605$. Since $Y_M$ is directly proportional to the amplitude of the pressure profile, and the overall tSZ power spectrum amplitude is proportional to the square of the amplitude pressure profile, it is a simple exercise to re-scale the tSZ power spectrum based on a given observational determined value of $Y_M$. 

Using this as a calibration would sidestep the need for any model calibration on simulations, but would add an additional error of 32\% to the tSZ power spectrum. Due to the large error we chose not to pursue this path in this paper, but in the future this could be an interesting alternative to the cluster abundance method, where weak lensing is used to calibrate the cluster masses. In tSZ power spectrum approach we would not need to count clusters, or understand the completeness, nor would we need to have cluster redshifts, making tSZ power spectrum approach a lot simpler. 

\subsubsection{Effects of AGN Physics}
\label{sec:AGN}
Various hydrodynamical simulations ~\cite[]{schaller2015effect} show large-scale expulsion of gas to the outer reaches halo, an effect that is stronger in low-mass halos due to their smaller gravitational potentials. 


We model the effect of baryonic feedback within the halo by the relative gas fraction first introduced in \cite[]{mohammed2014baryonic}, which appears as an overall normalization to the original gas pressure $\rho_0$ and therefore the Fourier transformed pressure profile $\tilde{y}_l$,
\begin{equation} f_{\textrm{gas}}(M_{halo},M_{crit}) = \frac{1}{1+\left(\frac{M_{crit}}{M_{halo}}\right)^2}.
\end{equation}

In the limit that there is no feedback, $M_{crit} \rightarrow 0$, the gas fraction function goes to one. 
While feedback effects can expulse the gas out to the virial radius 
and possibly beyond, the gas is not destroyed, and its thermal 
content is also unlikely to be changed, as also suggested by the 
weak lensing scaling 
relations of \cite[]{wang2016weak}.
We model this by enforcing integrated pressure conservation by exchanging the reduced pressure for a wide Gaussian profile,
\begin{equation} y_l^\textrm{new}(x, r_\textrm{vir}) =f_\textrm{gas} y^\textrm{0}_l (x) + (1-f_\textrm{gas})y_l^\textrm{feedback}(x, r_\textrm{vir}), \end{equation}
with a Gaussian profile of the form

\begin{equation} y_l^\textrm{feedback}(x, r_\textrm{vir}) = A(r_\textrm{vir})e^{-x^2/2(4r_\textrm{vir})^2}. \end{equation}

Here $A(r_\textrm{rvir})$ is a normalization coefficient calculated to preserve the overall integrated pressure. The spread of the Gaussian profile ($4r_\textrm{rvir}$) will affect the power suppression of 
large scales versus small scales. While this is admittedly a rather simplified approach, we have chosen the parameter such that it roughly quantitatively agrees with the AGN feedback results of \cite[]{mccarthy2014thermal}.

\begin{figure}
    \includegraphics[width=0.5\textwidth]{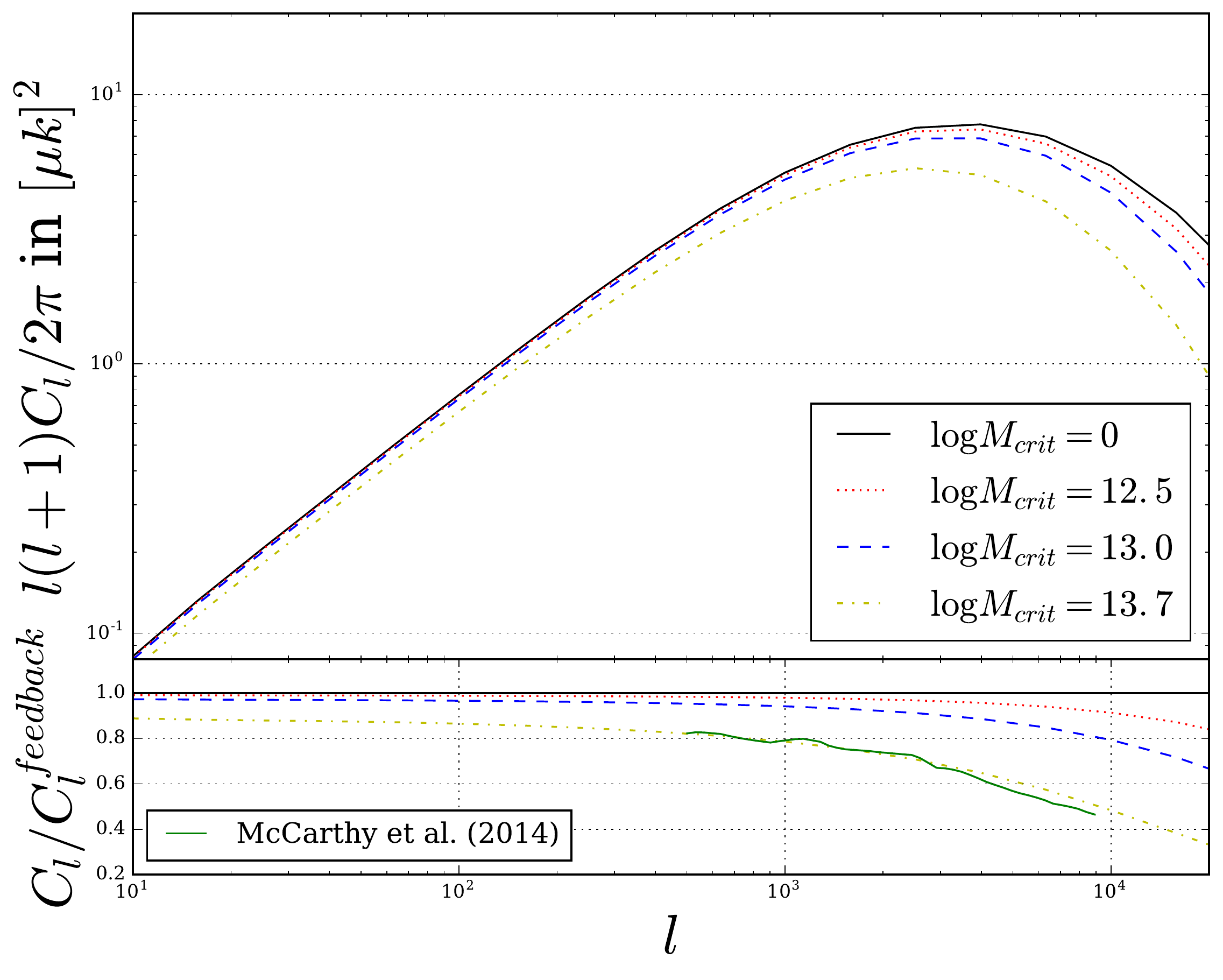}
    \caption{We show the effect of varying the $M_{crit}$ parameter on the power spectra at 146 GHz. We also plot the ratio of the power spectra relative to no feedback case at the bottom. Overall the feedback effects are not completely degenerate with amplitude $\sigma_8$, the parameter of interest, if sufficiently large range of scales is used in the analysis. At low $l$ the feedback effects are small. We do not show the full power spectrum of \protect\cite{mccarthy2014thermal} as their simulation volume is too small for this comparison to be meaningful, however we show the ratio of their low (AGN80) and high (AGN85) AGN feedback model to show the good agreement at high $l$ where AGN effects are most pronounced.}
    \label{fig2}
\end{figure}

The effects of varying the $M_\textrm{crit}$ parameter are shown in figure \ref{fig2}. We see power suppression at small angular scales ($l>3000$), as contributions from the low mass halos to the 1-halo term are suppressed, while effects at low $l$ are smaller, since large scales see most of the cluster gas even for very puffed-up halos. These same effects appear in tSZ power-spectra generated from simulations which include such feedback effects~\cite[]{mccarthy2014thermal}. 

\subsubsection{Covariance of the tSZ Power Spectrum}
\label{sec:AGN}

The tSZ power spectrum is dominated by the 1-halo term
. 
Accurate error calculations will rely on a non-diagonal covariance matrix with contributions from the full connected trispectrum, $T_{ll'}$, in addition 
to the gaussian (disconnected) diagonal term determined by $C_l^2$'s,
\begin{equation}M_{ll'} = \frac{1}{4 \pi f_{sky}} \left( \frac{4 \pi C_l^2}{l+1/2} \delta_{ll'} + T_{ll'}\right), \end{equation}
where $f_{sky}$ is the sky fraction of the observations used.
Here $C_l$ is the total tSZ power spectrum including the 
noise and systematics error (which we assume to be uncorrelated, an assumption that likely breaks down for 
foreground separation part). We follow ~\cite[]{cooray2001non,komatsu2002sunyaev} obtaining the dominant term for the trispectrum in the halo model as
\begin{equation}T_{ll'} \approx  \int_{z_{low}}^{z_{max}} dz \frac{d^2V}{d\Omega dz} \int_{M_{min}}^{M_{max}} dM \frac{dn(M,z)}{dM} |\tilde{y}_l(M,z)|^2 |\tilde{y}_{l'}(M,z)|^2 .\end{equation}
Note that while the disconnected term is diagonal, 
the connected term $T_{ll'}$ is not and it is indeed very 
strongly correlating the bins. This term dominates the 
covariance at lower $l$, where the disconnected term is 
small. 

In addition to this we would also need to include a beat-coupling (super-sample variance) term, that is determined by the 
variance of the effective volume contributing to a given $l$. 
This term gas been computed by \cite[]{schaan2014joint}, and has been 
shown to be subdominant compared to the connected trispectrum term.\cite[]{hill2013cosmology} \cite[]{chPrivate} We will hence ignore it in our analysis. 

To these theoretical covariance terms we add the noise and 
foreground variance from Planck data. The latter is dominated by the imperfect separation between tSZ and cosmic infrared 
background (CIB). We take the numbers 
as given in ~\cite[]{aghanim2015planck}. We assume these errors 
are uncorrelated, which is probably not completely valid for 
CIB component separation, however we find that this doesn't significantly effect the analysis.
Like the tSZ power spectra, the trispectrum is sensitive to the cosmology (particularly $\sigma_8$ and $\Omega_m$) and we vary it with the cosmology in our analysis.


\section{Analysis and Results}
\label{sec:Analysis}

In this section we describe a procedure for constraining cosmological parameters while using this new $f_{gas}$ parametrization. We use this to particularly constrain $\sigma_8$, however this procedure generalizes to other cosmological parameters.

We begin by writing a likelihood function of the tSZ power spectrum with relation to a combined Planck + ACT dataset.

\begin{equation}\log{\mathcal{L}(M_{crit},\sigma_8)} \propto  \sum_{l \leq l'} (\hat{C}_l - C_l) (M^{-1})_{ll'} (\hat{C}_{l'} - C_{l'})\end{equation}

We use a flat prior on $\sigma_8$ between 0.70 and 0.9, covering the ranges of parameters found from a number of other CMB-based surveys. Motivated by the discussion in \cite{mohammed2014baryonic}, we explicitly exclude unphysical values of $\log{M_{crit}}>14.5$. Values $\log{M_{crit}}<11.5$ would not affect the tSZ power spectrum as such light halos do not significantly contribute except at extraordinarily high $l$ and are below our mass integration range. 

\begin{figure}
    \includegraphics[width=240px]{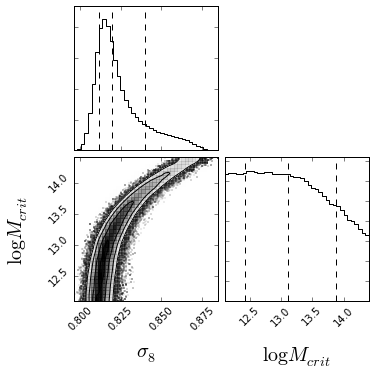}
    \caption{The result of our MCMC calculation with lines indicating the one standard deviation spread at the 16th, 50th, and 84th percentiles.}
    \label{fig_mcmc}
\end{figure}

\begin{figure*}
    \includegraphics[width=1.0\textwidth]{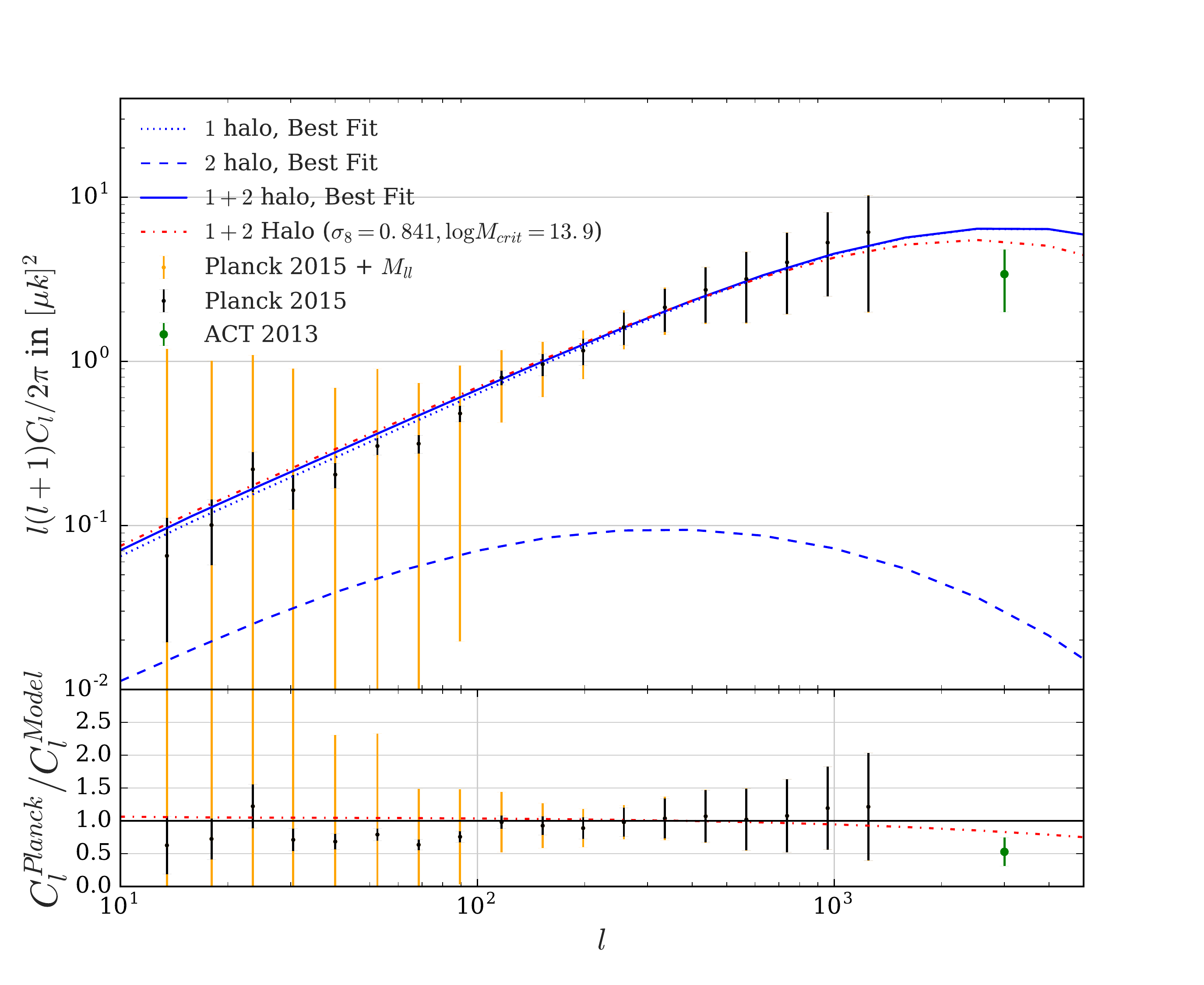}
    \caption{The best fit model ($\sigma_8=0.820$, $\log{M_{crit}}=13.11$) at 146 GHz. Data and errors shown are those quoted from Planck 2015 \protect\cite[]{aghanim2015planck} and ACT 2013 \protect\cite{hill2014atacama}. We have also included the power spectrum $\sigma_8 = 0.841$, the one standard deviation above our fit, and at the corresponding best fit value of $\log{M_{crit}}=13.9$. This curve reflects the effect of an increased weight to the high $l$ data from SPT.}
    \label{fig_fit}
\end{figure*}

We use a Markov Chain Monte Carlo routine to explore the likelihood space and find likelihood distributions for our two parameters of interest, shown in figure \ref{fig_mcmc}. We quote the 50th percentile value as our best fit and the 16th and 84th as our associated errors. The power-spectrum with the best fit values is shown in figure \ref{fig_fit}.

Our fit is dominated by values at intermediate $l$ (between $l=90$ and $l=350$), as those at higher $l$ have large errors dominated by systematic uncertainty (both due to CIB and due to AGN feedback modeling), while those at lower l are dominated by the connected trispectrum, originating from the small effective volume of tSZ power spectrum at low $l$. In figure \ref{fig_fit} we include error bars from noise and foreground uncertainties, as well as a trispectrum error term of the form:

\begin{equation}\sigma_\textrm{tri}=\frac{l(l+1)}{2\pi} \left( \frac{T_{ll} }{4 \pi f_\textrm{sky}} \right)\end{equation}

A summary of data values used, associated errors, and residuals is found in table \ref{table1}.
We see that the region from $100<l<350$ has errors that 
are small enough to provide meaningful constraints to cosmological parameters, with the minimum combined error 
at $l \sim 150$ with the relative error of 26\%. 
Using $\sigma_8^8$ scaling this gives a 3\% error 
on $\sigma_8$ from a single bin. The bins are strongly 
correlated due to the connected 
trispectrum and combining them provides some 
modest gain over a single bin analysis,
\begin{equation}
\sigma_8 =0.820^{+0.021}_{-0.009}\left(\Omega_m/0.31\right)^{0.4}
\end{equation}
This result assumes Planck values of other 
cosmological parameters. Note that the errors are 
somewhat asymmetric: this is because a high $\sigma_8$ 
can be compensated by a high value of $M_{crit}$, while 
lower values of $M_{crit}$ have no impact on the tSZ 
power spectrum. Our derived amplitude is somewhat higher than a similar analysis presented in \cite{hill2014detection} based on Planck 2013 tSZ power spectrum. 
This is because the Planck 2015 tSZ power spectrum is significantly higher than Planck 2013. We suspect this change 
is mostly due to improvements in the foreground separation 
method, specially CIB separation.  While our analysis assumes that these foreground errors are uncorrelated, changing those assumptions to allow for more strongly correlated foreground errors doesn't change the results in a statistically significant way.

\begin{table*}
\caption{Planck + ACT Data and our Best Fit. Note that $\sigma_\textrm{tri}$ is only the diagonal part, with 
substantial off-diagonal terms we do not show here. We have 
assumed $\sigma_\textrm{stat + fg}$ component 
to be diagonal, since only diagonal part is provided by Planck team. All values are at 146 GHz (where $g_\nu^2 = 1$) and all units are in $[\mu K]^2$.}
\label{my-label}
\begin{tabular}{l|c|c|c|c|c|c}
\hline
\hline
$l_\textrm{eff}$ & $l(l+1)C^\textrm{data}_l/2\pi$ & $\sigma_\textrm{stat + fg}$ & $\sigma_\textrm{tri}$ & $\sigma_\textrm{all}$ & $l(l+1)C^\textrm{best-fit}_l/2\pi$ & Res \\
\hline
\hline
13.5   & 0.065 & 0.0458 & 1.1220 & 1.1230 & 0.1040 & -0.0388 \\
18.0   & 0.101 & 0.0432 & 0.9071 & 0.9081 & 0.1388 & -0.0382 \\
23.5   & 0.220 & 0.0600 & 0.8693 & 0.8713 & 0.1798 & 0.0399  \\
30.5   & 0.164 & 0.0389 & 0.7377 & 0.7388 & 0.2304 & -0.0665 \\
40.0   & 0.204 & 0.0353 & 0.4830 & 0.4843 & 0.2985 & -0.0944 \\
52.5   & 0.305 & 0.0356 & 0.5924 & 0.5935 & 0.3857 & -0.0809 \\
68.5   & 0.315 & 0.0399 & 0.4198 & 0.4217 & 0.4957 & -0.1809 \\
89.5   & 0.481 & 0.0535 & 0.4584 & 0.4615 & 0.6357 & -0.1546 \\
117.0  & 0.797 & 0.0800 & 0.3644 & 0.3730 & 0.8133 & -0.0164 \\
152.5  & 0.961 & 0.1468 & 0.3226 & 0.3545 & 1.0379 & -0.0772 \\
198.0  & 1.162 & 0.2134 & 0.3161 & 0.3814 & 1.3070 & -0.1452 \\
257.5  & 1.616 & 0.3611 & 0.2353 & 0.4310 & 1.6509 & -0.0348 \\
335.5  & 2.134 & 0.6212 & 0.2787 & 0.6808 & 2.0568 & 0.0769  \\
436.5  & 2.726 & 1.0079 & 0.2307 & 1.0340 & 2.5484 & 0.1774  \\
567.5  & 3.177 & 1.4552 & 0.2503 & 1.4766 & 3.1143 & 0.0628  \\
738.0  & 4.013 & 2.0707 & 0.2346 & 2.0839 & 3.7307 & 0.2828  \\
959.0  & 5.295 & 2.7985 & 0.2252 & 2.8076 & 4.4373 & 0.8579  \\
1247.0 & 6.128 & 4.1254 & 0.4374 & 4.1486 & 5.0495 & 1.0790 \\
3000.0 & 3.4 & 1.4 & 0.8291 & 1.6 & 6.4365 & -3.0 \\
\hline
\end{tabular}
\label{table1}
\footnotetext{Value and error bars from ACT 2013.}
\end{table*}
\section{Discussion and Outlook}
\label{sec:discussion}

In this paper we present a tSZ power spectrum cosmology analysis, where we extend previous work by accounting for feedback in clusters and including connected trispectrum in the analysis. We introduce a one-parameter model for feedback within galactic clusters which, when marginalized over, provides more realistic constraints on cosmological parameters. Our model is relatively simple, but also does not rely on any detailed understanding of gas dynamics or simulated results. A key assumption of our model is that 
while astrophysical processes within the cluster can push the 
gas around, possibly all the way to the virial radius, suppressing 
small scale clustering, its total thermal content does not change, 
guaranteeing that large scale clustering is unchanged. 
Maximizing likelihood with respect to this model we find we an updated constraint for $\sigma_8 =0.820^{+0.021}_{-0.009}\left(\Omega_m/0.31\right)^{0.4}$ when using our combined Planck/ACT data set, assuming Planck cosmology values for other parameters. Our results are consistent with Planck's overall normalization of $\sigma_8^{\textrm{Planck}}= 0.8159 \pm 0.0086$~\cite[]{planck2015cosmo}. 

Planck SZ analysis has been argued to be supporting low amplitude, $\sigma_8 = 0.78 \pm 0.02$ ~\cite[]{aghanim2015planck}, but we find no evidence of this in our tSZ power spectrum analysis. If anything, our results suggest a normalization that is even higher than that of Planck cosmology, which in itself is relatively high. Our analysis differs from that done in Planck in two substantial ways; we have a full treatment of the trispectrum which substantially reduces the weight of low-$l$ data which would otherwise support a lower value of $\sigma_8$ and the use of marginalization over feedback parameters which effectively reduces the weight of high-$l$ data. 
Our normalization is also somewhat higher than the values given by ACT-SZ analysis $\sigma_8^{\textrm{ACT-SZ}} = 0.793 \pm 0.018$~\cite[]{hill2014atacama}. Our results also differ from those found in \cite[]{hill2014detection}  primarily due to a substantial amplitude shift between Planck 2013 and Planck 2015 tSZ Compton parameter map.

It should be noted that we do not include data from the South Pole Telescope (SPT)\cite[]{george2015measurement}. There has been discussion in the literature about possible challenges facing this measurement due to the limited frequency channels compared to Planck, creating difficulty distinguishing tSZ signal from the primary CMB and other extra-galactic sources. This difficultly is further compounded by the fact that at $l=3000$ the tSZ signal is sub dominant to the foregrounds\cite[]{dolag2015sz}. While these same challenges face ACT's tSZ analysis, we use ACT's value due to their higher error bars which we view as more reflective of the current uncertainty in this difficult measurement. Like previous work, our model has more small scale power than predicted by SPT (or ACT) and it is likely that additional data at $l>2000$ will help resolve this tension.

We have identified $100<l<350$ as the sweet spot where the current errors and modeling uncertainties are minimized; at $l<100$ the amplitude has significant trispectrum error and at $l>350$ the amplitude has significant error from AGN modeling and foregrounds. CIB contamination is one of the dominant sources of error, and if the foreground separation can be improved one may be able to push the analysis to higher $l$, where connected trispectrum term become less important. However, pushing to higher $l$ would also require improved modeling of feedback effects on the pressure profile, which is not 
necessary for low $l$ used in this paper. 

Similarly, one might be able to expand the "sweet spot" to lower $l$ by masking nearby clusters. This has been shown (\cite[]{shaw2009sharpening}, \cite[]{hill2013cosmology}) as a possible technique to reduce the trispectrum error, since large nearby clusters contribute significantly to the signal at low $l$, but contain little volume and hence are subject to significant sampling variance and Poisson fluctuations. 

Going forward it will be useful to further extend our model based on a more nuanced understanding of the gas physics involved. Our model is fairly simplistic in how we treat the expelled gas pressure distribution. It is possible that with new tSZ, kSZ, and X-ray studies understanding the pressure profile at the outer edges of clusters would be improved. Similarly, our flat prior for $\log{M_{crit}}$ is simple and better independent constraints of this parameter would allow similarly better constraints of $\sigma_8$. 

As new generations of CMB experiments start collecting data at higher spatial resolutions, one would like to be able to further improve cosmological constraints from the tSZ power spectrum.
Our analysis suggests this may in principle be possible, but it will not be easy, and will require improvements in our understanding of 
the cluster pressure profiles on scales at and beyond the virial radius. 


\section*{Acknowledgments}
We are thankful to Eiichiro Komatsu for his publicly available cosmology routines. We thank Nick Battaglia, Colin Hill, Chirag Modi, and Oliver Zahn for helpful conversation. BH is supported by NSF Graduate Research Fellowship, award number DGE 1106400. We acknowledge support of NASA grant NNX15AL17G.

This research used resources of the National Energy Research Scientific Computing Center, a DOE Office of Science User Facility supported by the Office of Science of the U.S. Department of Energy under Contract No. DE-AC02-05CH11231.

\bibliographystyle{mnras}
\bibliography{SZpaperb}

\begin{thebibliography}{}
\makeatletter
\relax
\def\mn@urlcharsother{\let\do\@makeother \do\$\do\&\do\#\do\^\do\_\do\%\do\~}
\def\mn@doi{\begingroup\mn@urlcharsother \@ifnextchar [ {\mn@doi@}
  {\mn@doi@[]}}
\def\mn@doi@[#1]#2{\def\@tempa{#1}\ifx\@tempa\@empty \href
  {http://dx.doi.org/#2} {doi:#2}\else \href {http://dx.doi.org/#2} {#1}\fi
  \endgroup}
\def\mn@eprint#1#2{\mn@eprint@#1:#2::\@nil}
\def\mn@eprint@arXiv#1{\href {http://arxiv.org/abs/#1} {{\tt arXiv:#1}}}
\def\mn@eprint@dblp#1{\href {http://dblp.uni-trier.de/rec/bibtex/#1.xml}
  {dblp:#1}}
\def\mn@eprint@#1:#2:#3:#4\@nil{\def\@tempa {#1}\def\@tempb {#2}\def\@tempc
  {#3}\ifx \@tempc \@empty \let \@tempc \@tempb \let \@tempb \@tempa \fi \ifx
  \@tempb \@empty \def\@tempb {arXiv}\fi \@ifundefined
  {mn@eprint@\@tempb}{\@tempb:\@tempc}{\expandafter \expandafter \csname
  mn@eprint@\@tempb\endcsname \expandafter{\@tempc}}}

\bibitem[\protect\citeauthoryear{Arnaud, Pratt, Piffaretti, B{\"o}hringer,
  Croston  \& Pointecouteau}{Arnaud et~al.}{2010}]{arnaud2010universal}
Arnaud M.,  Pratt G.,  Piffaretti R.,  B{\"o}hringer H.,  Croston J.,
  Pointecouteau E.,  2010, Astronomy \& Astrophysics, 517, A92

\bibitem[\protect\citeauthoryear{Battaglia, Bond, Pfrommer, Sievers  \&
  Sijacki}{Battaglia et~al.}{2010}]{battaglia2010simulations}
Battaglia N.,  Bond J.,  Pfrommer C.,  Sievers J.,   Sijacki D.,  2010, The
  Astrophysical Journal, 725, 91

\bibitem[\protect\citeauthoryear{Battaglia, Bond, Pfrommer  \&
  Sievers}{Battaglia et~al.}{2012}]{battaglia2012cluster}
Battaglia N.,  Bond J.,  Pfrommer C.,   Sievers J.,  2012, The Astrophysical
  Journal, 758, 75

\bibitem[\protect\citeauthoryear{Cooray}{Cooray}{2001}]{cooray2001non}
Cooray A.,  2001, Physical Review D, 64, 063514

\bibitem[\protect\citeauthoryear{Dolag, Komatsu  \& Sunyaev}{Dolag
  et~al.}{2015}]{dolag2015sz}
Dolag K.,  Komatsu E.,   Sunyaev R.,  2015, arXiv preprint arXiv:1509.05134

\bibitem[\protect\citeauthoryear{Duffy, Schaye, Kay  \& Dalla~Vecchia}{Duffy
  et~al.}{2008}]{duffy2008dark}
Duffy A.~R.,  Schaye J.,  Kay S.~T.,   Dalla~Vecchia C.,  2008, Monthly Notices
  of the Royal Astronomical Society: Letters, 390, L64

\bibitem[\protect\citeauthoryear{George et~al.,}{George
  et~al.}{2015}]{george2015measurement}
George E.,  et~al., 2015, The Astrophysical Journal, 799, 177

\bibitem[\protect\citeauthoryear{Greco, Hill, Spergel  \& Battaglia}{Greco
  et~al.}{2015}]{greco2015stacked}
Greco J.~P.,  Hill J.~C.,  Spergel D.~N.,   Battaglia N.,  2015, The
  Astrophysical Journal, 808, 151

\bibitem[\protect\citeauthoryear{Hill \& Pajer}{Hill \&
  Pajer}{2013}]{hill2013cosmology}
Hill J.~C.,  Pajer E.,  2013, Physical Review D, 88, 063526

\bibitem[\protect\citeauthoryear{Hill \& Schaan}{Hill \&
  Schaan}{2016}]{chPrivate}
Hill C.,  Schaan E.,  2016, Private Communication

\bibitem[\protect\citeauthoryear{Hill \& Spergel}{Hill \&
  Spergel}{2014}]{hill2014detection}
Hill J.~C.,  Spergel D.~N.,  2014, Journal of Cosmology and Astroparticle
  Physics, 2014, 030

\bibitem[\protect\citeauthoryear{Hill et~al.,}{Hill
  et~al.}{2014}]{hill2014atacama}
Hill J.~C.,  et~al., 2014, arXiv preprint arXiv:1411.8004

\bibitem[\protect\citeauthoryear{Komatsu \& Kitayama}{Komatsu \&
  Kitayama}{1999}]{komatsu1999sunyaev}
Komatsu E.,  Kitayama T.,  1999, The Astrophysical Journal Letters, 526, L1

\bibitem[\protect\citeauthoryear{Komatsu \& Seljak}{Komatsu \&
  Seljak}{2002}]{komatsu2002sunyaev}
Komatsu E.,  Seljak U.,  2002, Monthly Notices of the Royal Astronomical
  Society, 336, 1256

\bibitem[\protect\citeauthoryear{McCarthy, Le~Brun, Schaye  \& Holder}{McCarthy
  et~al.}{2014}]{mccarthy2014thermal}
McCarthy I.~G.,  Le~Brun A.,  Schaye J.,   Holder G.,  2014, Monthly Notices of
  the Royal Astronomical Society, 440, 3645

\bibitem[\protect\citeauthoryear{Mohammed, Martizzi, Teyssier  \&
  Amara}{Mohammed et~al.}{2014}]{mohammed2014baryonic}
Mohammed I.,  Martizzi D.,  Teyssier R.,   Amara A.,  2014, arXiv preprint
  arXiv:1410.6826

\bibitem[\protect\citeauthoryear{Nozawa, Itoh, Suda  \& Ohhata}{Nozawa
  et~al.}{2005}]{nozawa2005relativistic}
Nozawa S.,  Itoh N.,  Suda Y.,   Ohhata Y.,  2005, arXiv preprint
  astro-ph/0507466

\bibitem[\protect\citeauthoryear{{Planck Collaboration} et~al.,}{{Planck
  Collaboration} et~al.}{2013}]{2013A&A...557A..52P}
{Planck Collaboration} et~al., 2013, \mn@doi [\aap]
  {10.1051/0004-6361/201220941}, 557, A52

\bibitem[\protect\citeauthoryear{{Planck Collaboration} et~al.}{{Planck
  Collaboration} et~al.}{2015}]{planck2015cosmo}
{Planck Collaboration} et~al., 2015, arXiv preprint arXiv:1502.01589

\bibitem[\protect\citeauthoryear{Planck et~al.}{Planck
  et~al.}{2015}]{aghanim2015planck}
Planck C.,  et~al., 2015, arXiv preprint arXiv:1502.01596

\bibitem[\protect\citeauthoryear{Pontzen \& Governato}{Pontzen \&
  Governato}{2012}]{pontzen2012supernova}
Pontzen A.,  Governato F.,  2012, Monthly Notices of the Royal Astronomical
  Society, 421, 3464

\bibitem[\protect\citeauthoryear{Read \& Gilmore}{Read \&
  Gilmore}{2005}]{read2005mass}
Read J.,  Gilmore G.,  2005, Monthly Notices of the Royal Astronomical Society,
  356, 107

\bibitem[\protect\citeauthoryear{Reese, Carlstrom, Joy, Mohr, Grego  \&
  Holzapfel}{Reese et~al.}{2002}]{reese2002determining}
Reese E.~D.,  Carlstrom J.~E.,  Joy M.,  Mohr J.~J.,  Grego L.,   Holzapfel
  W.~L.,  2002, The Astrophysical Journal, 581, 53

\bibitem[\protect\citeauthoryear{Schaan, Takada  \& Spergel}{Schaan
  et~al.}{2014}]{schaan2014joint}
Schaan E.,  Takada M.,   Spergel D.~N.,  2014, Physical Review D, 90, 123523

\bibitem[\protect\citeauthoryear{Schaller et~al.,}{Schaller
  et~al.}{2015}]{schaller2015effect}
Schaller M.,  et~al., 2015, Monthly Notices of the Royal Astronomical Society,
  452, 343

\bibitem[\protect\citeauthoryear{Shaw, Zahn, Holder  \& Dor{\'e}}{Shaw
  et~al.}{2009}]{shaw2009sharpening}
Shaw L.~D.,  Zahn O.,  Holder G.~P.,   Dor{\'e} O.,  2009, The Astrophysical
  Journal, 702, 368

\bibitem[\protect\citeauthoryear{Shaw, Nagai, Bhattacharya  \& Lau}{Shaw
  et~al.}{2010}]{shaw2010impact}
Shaw L.~D.,  Nagai D.,  Bhattacharya S.,   Lau E.~T.,  2010, The Astrophysical
  Journal, 725, 1452

\bibitem[\protect\citeauthoryear{Story et~al.,}{Story
  et~al.}{2013}]{story2013measurement}
Story K.,  et~al., 2013, The Astrophysical Journal, 779, 86

\bibitem[\protect\citeauthoryear{Sunyaev \& Zeldovich}{Sunyaev \&
  Zeldovich}{1970}]{sunyaev1970small}
Sunyaev R.~A.,  Zeldovich Y.~B.,  1970, Astrophysics and Space Science, 7, 3

\bibitem[\protect\citeauthoryear{Tinker, Kravtsov, Klypin, Abazajian, Warren,
  Yepes, Gottl{\"o}ber  \& Holz}{Tinker et~al.}{2008}]{tinker2008mf}
Tinker J.,  Kravtsov A.~V.,  Klypin A.,  Abazajian K.,  Warren M.,  Yepes G.,
  Gottl{\"o}ber S.,   Holz D.~E.,  2008, The Astrophysical Journal, 688, 709

\bibitem[\protect\citeauthoryear{Tinker, Robertson, Kravtsov, Klypin, Warren,
  Yepes  \& Gottl{\"o}ber}{Tinker et~al.}{2010}]{tinker2010bias}
Tinker J.~L.,  Robertson B.~E.,  Kravtsov A.~V.,  Klypin A.,  Warren M.~S.,
  Yepes G.,   Gottl{\"o}ber S.,  2010, The Astrophysical Journal, 724, 878

\bibitem[\protect\citeauthoryear{Trac, Bode  \& Ostriker}{Trac
  et~al.}{2011}]{trac2011templates}
Trac H.,  Bode P.,   Ostriker J.~P.,  2011, The Astrophysical Journal, 727, 94

\bibitem[\protect\citeauthoryear{Wang, White, Mandelbaum, Henriques, Anderson
  \& Han}{Wang et~al.}{2016}]{wang2016weak}
Wang W.,  White S.~D.,  Mandelbaum R.,  Henriques B.,  Anderson M.~E.,   Han
  J.,  2016, Monthly Notices of the Royal Astronomical Society, 456, 2301

\makeatother
\end{thebibliography}

\end{document}